\documentclass[a4paper,12pt]{article}
\usepackage{graphicx,amssymb,color}
\usepackage{epsfig}
\usepackage{color}

\begin{document}

\vspace*{3mm}

\centerline{\LARGE \bf  On a quarantine model of coronavirus infection}

\bigskip

\centerline{\LARGE \bf  and data analysis}

\vspace*{1cm}

\centerline{\bf Vitaly Volpert$^{1,2,3,*}$, Malay Banerjee$^4$, Sergei Petrovskii$^5$}

\vspace*{0.5cm}

\centerline{$^1$
Institut Camille Jordan, UMR 5208 CNRS, University Lyon 1, 69622 Villeurbanne, France}

\centerline{$^2$
INRIA
Team Dracula, INRIA Lyon La Doua, 69603 Villeurbanne, France}

\centerline{$^3$
Peoples’ Friendship University of Russia (RUDN University)}
\centerline{ 6 Miklukho-Maklaya St, Moscow, 117198, Russia}

\centerline{$^4$ Department of Mathematics \& Statistics, IIT Kanpur, Kanpur - 208016, India}

\centerline{$^5$
School of Mathematics \& Actuarial Science, University of Leicester, LE1 7RH, UK}

\vspace*{1cm}

\noindent
{\bf Abstract.}
Attempts to curb the spread of coronavirus by introducing strict quarantine measures apparently have different effect in different countries: while the number of new cases has reportedly decreased in China and South Korea, it still exhibit significant growth in Italy and other countries across Europe. In this brief note, we endeavour to assess the efficiency of quarantine measures by means of mathematical modelling.
Instead of the classical SIR model, we introduce a new model of infection progression under the assumption that all infected individual are isolated after the incubation period in such a way that they cannot infect other people. Disease progression in this model is determined by the basic reproduction number $\mathcal{R}_0$ (the number of newly infected individuals during the incubation period), which is different compared to that for the standard SIR model. If $\mathcal{R}_0 >1$, then the number of latently infected individuals exponentially grows. However, if $\mathcal{R}_0 <1$ (e.g.~due to quarantine measures and contact restrictions imposed by public authorities), then the number of infected decays exponentially. We then consider the available data on the disease development in different countries to show that there are three possible patterns: growth dynamics, growth-decays dynamics, and patchy dynamics (growth-decay-growth). Analysis of the data in China and Korea shows that the peak of infection (maximum of daily cases) is reached about 10 days after the restricting measures are introduced. During this period of time, the growth rate of the total number of infected was gradually decreasing. However, the growth rate remains exponential in Italy. Arguably, it suggests that the introduced quarantine is not sufficient and stricter measures are needed.

\vspace*{5mm}


\vspace*{15mm}

\noindent
$(^*)~$Corresponding author. Email: volpert@math.univ-lyon1.fr

\vspace*{1cm}

\section{SIR model}

The classical suspeceptible-infected-recovered (SIR) model in epidemiology \cite{H,M} allows the determination of critical condition of disease development in the population irrespective of total population size over a short period of time. Among many versions of the model, let us consider the following simplest ODE system:

\begin{equation}\label{a1}
  \frac{dS}{dt} = - k IS ,
\end{equation}

\begin{equation}\label{a2}
  \frac{dI}{dt} =  k IS - \beta I - \sigma I ,
\end{equation}

\begin{equation}\label{a3}
  \frac{dR}{dt} =   \beta I,
\end{equation}
for the sizes of the susceptible sub-population $S$, infected $I$, and recovered $R$. The term $kIS$ describes the disease transmission rate due to the contacts between susceptible and infected individuals, $\beta I$ characterizes the rate of recovery of infected, and $\sigma I$ mortality due to infection only. It is assumed here that recovered individuals do not return to susceptible class, that is, recovered individuals have immunity against the disease; they cannot become infected again and cannot infect susceptibles either.

Analysis of this model is well-known \cite{H,M} and we will not discuss it here. Since the coronavirus epidemics is admittedly still in its early stage, let us only determine the condition of the disease progression at its beginning, i.e.~when the number of infected/recovered/dead is much less than the number of susceptible, and hence $S$ in the model above can be considered as constant, $S \approx S_0$. The equation (\ref{a2}) then turns into the following:

\begin{equation}\label{a4}
  \frac{dI}{dt} =  k IS_0 - \beta I - \sigma I .
\end{equation}
so that it splits off the system (\ref{a1}--\ref{a3}) and can be considered separately. It is an ordinary differential equation with constant coefficients whose solution can readily be found, $I(t) = I_0 e^{\mu t}$, where $I_0$ is the initial number of infected ($I_0\ge 1$), i.e.~at the time when the infection/disease was first detected. Substituting this expression into equation (\ref{a4}), we get

$$ \mu = (kS_0 - \beta - \sigma) = (\beta+\sigma) (\mathcal{R}_0 - 1) , $$
where the new parameter $\mathcal{R}_0=kS_0/(\beta+\sigma)$ is called the basic reproduction number. If $\mathcal{R}_0>1$, then the number of infected will grow, if $\mathcal{R}_0<1$ it will decay. Thus, $\mathcal{R}_0$ plays the key role in the epidemics development. In particular, one can change the course of the disease dynamics by changing $\mathcal{R}_0$. For instance, if measures are introduced that push the value of$\mathcal{R}_0$ below one (e.g.~by making $k$ to decrease), the exponential growth changes to exponential decay. Interestingly, this is exactly what has happened in South Korea (see Section \ref{sec:data}) after restrictions on daily life were introduced.

The classical SIR model therefore determines the condition of the disease development from the comparison of the disease transmission rate with the sum of the recover and death rates. In the other words, we compare the rates of adding and removal of infected. The model does not account for the incubation period of the disease that was shown to be important in the case of coronavirus spread: individuals can become infective before showing any symptoms. A model that allows for the incubation period will be considered in the next section.


\setcounter{equation}{0}

\section{Quarantine model}

\paragraph{Model.}

Exceptional measures adopted for the coronavirus infection suggest to introduce another model of infection development. We consider the sub-population of latently infected individuals who are already infected but do not show any symptom during the incubation period. When the incubation period is over, the disease manifests itself with its symptoms, and the individual is isolated in the quarantine where he/she cannot infect the others. Under these assumptions, instead of system (\ref{a1})-(\ref{a3}) we have

\begin{equation}\label{b1}
  \frac{dS}{dt} = - k I(t)S(t) ,
\end{equation}

\begin{equation}\label{b2}
  \frac{dI}{dt} =  k I(t)S(t) - k I(t-\tau)S(t-\tau) ,
\end{equation}
where $I$ is the sub-population of latently infected individuals, $\tau$ is the incubation period. The second term in the right-hand side of equation (\ref{b2}) corresponds to the individuals infected at time $t-\tau$. Their incubation period is finished at time $t$, and they are put to quarantine. As a result they can not speard infection anymore.

\paragraph{Solution.}

\begin{figure}[ht!]
\centerline{\includegraphics[scale=0.5]{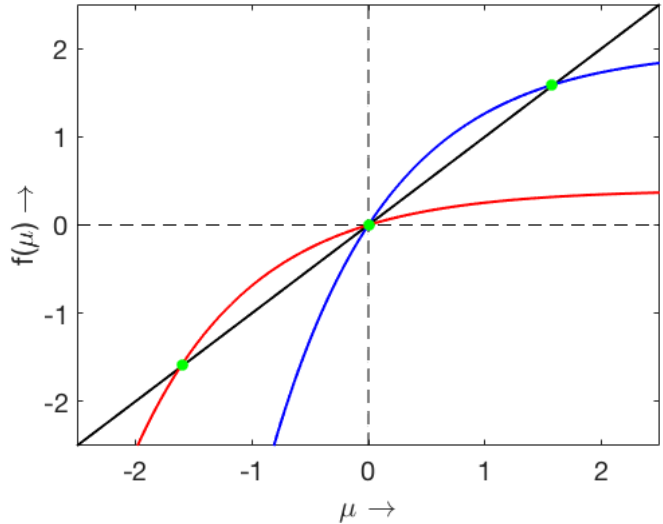}}
\caption{Graphical solution of equation (\ref{b4}), $\mu = f(\mu)$ with the values of parameters: $\tau=1$, $kS_0=2$ (blue curve) and $kS_0=0.4$ (red curve).}
\label{fig}
\end{figure}

As before, we approximate $S(t) \approx S_0$, $S(t-\tau) \approx S_0$ as $\tau$ is not large, and replace equation (\ref{b2}) by the equation

\begin{equation}\label{b3}
  \frac{dI}{dt} =  k I(t)S_0 - k I(t-\tau) S_0 .
\end{equation}
Substituting $I(t) = I_0 e^{\mu t}$ in above equation, we get

\begin{equation}\label{b4}
  \mu = kS_0 (1 - e^{-\mu \tau} ) .
\end{equation}
This equation has solution $\mu=0$ for all values of parameters. Besides this solution, it can have a positive or a negative solution $\mu$. The function $f(\mu) = kS_0 (1 - e^{-\mu \tau} )$ is an increasing function with the asymptotic limit $kS_0$ at infinity. The sign of the solution is determined by the derivative $f'(0)$. If $f'(0)>1$, then there is a positive solution (Figure \ref{fig}), if $f'(0)<1$, the solution is negative. In terms of the basic reproduction number $\mathcal{R}_0=kS_0\tau$, the condition is similar as for the SIR model. However, the meaning of this parameter is different. It characterizes the total infection rate during the incubation period and not its ratio with the rates of recover and death.

With the notation $\hat \mu = \mu/(kS_0)$, we can write equation (\ref{b4}) as follows: 

\begin{equation}\label{b15}
  \hat \mu = 1 - e^{-\mathcal{R}_0 \hat \mu}
\end{equation}
Its solution depends on the single parameter $\mathcal{R}_0$.

\paragraph{Public health.}

In the absence of vaccination and lack of effective treatment, the only way to influence the disease development is to act on the basic reproduction number $\mathcal{R}_0 = kS_0 \tau$, that is to decrease the value of the parameter $k$ quantifying the disease transmission rate by the infected individuals (among which the symptoms are not prominent yet). Note $S_0$ is a characteristic of the given population and $\tau$ is the property of the disease, therefore neither of them is controllable. However, parameter $k$ depends on the human mobility and the social distance and can be changed. If in the beginning of the disease development $k > 1/(S_0 \tau)$, that is each infected individual contaminate in average more than one susceptible individual during the incubation period, then the number of newly infected individuals will grow exponentially. If, at some moment of time, measures restricting potential contacts between infected and susceptible individuals are introduced - the quarantine, then $k$ can become less than the critical value, and the number of newly infected individuals will start decaying exponentially, thus resulting in a transition between the growing and decaying exponentials.

\section{Data}\label{sec:data}

 We used the data from \cite{W} showing the total number of infected individuals and daily cases in different countries and worldwide. Across the world, daily cases clearly show two-mode dynamics such as in China (growth-decay) and Europe (growth). There are periods of exponential growth, decay and re-growth. The origin of the outbreak on February 12 (in China) is not clear yet. It may be related to the method of data collection.
Daily reported cases curves in China and South Korea correspond to the growth-decay dynamics.
The daily new cases curves for France, Germany, Italy, Spain, USA correspond to the growth mode.

\begin{figure}[ht!]
\centerline{\includegraphics[scale=0.45]{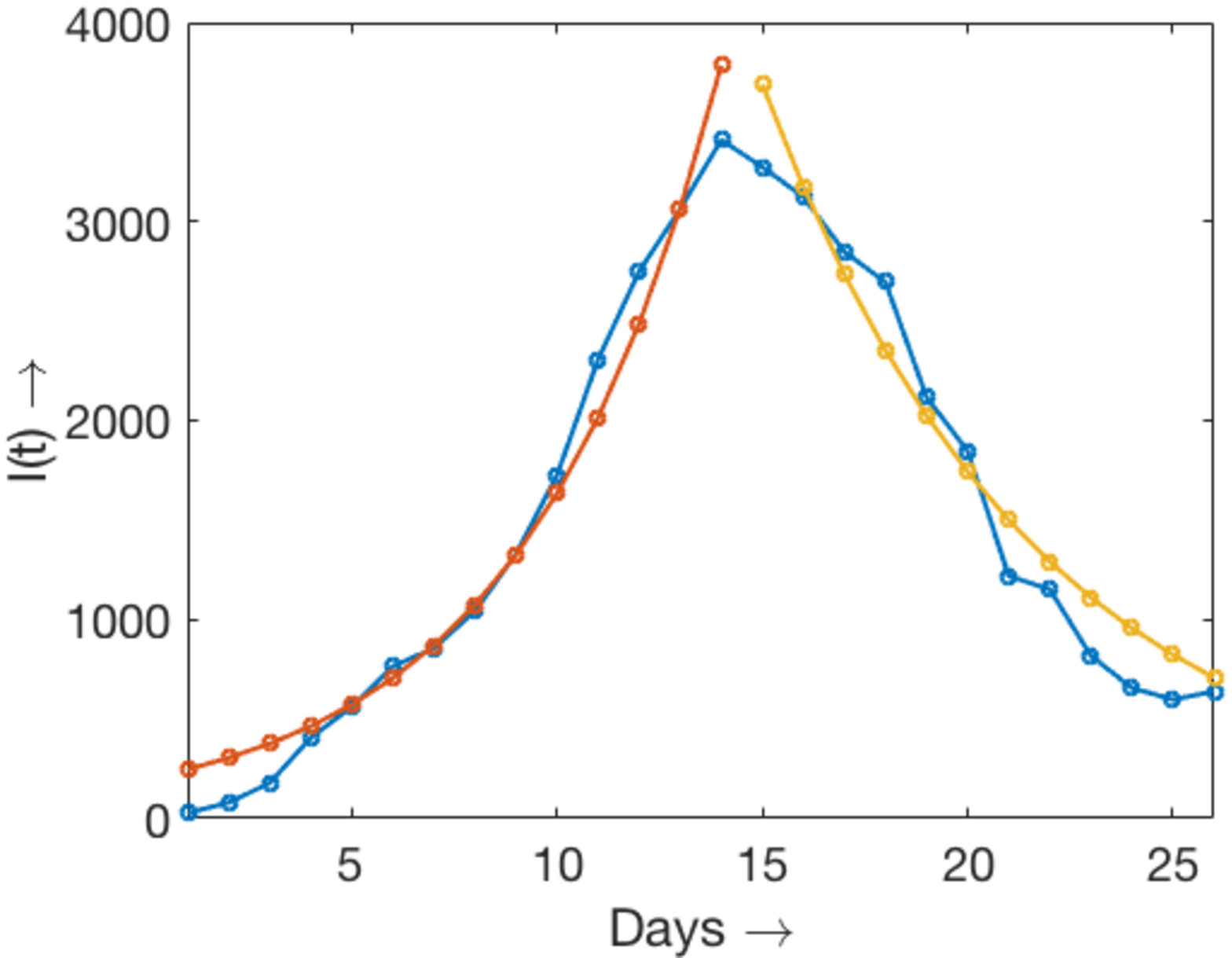}
\includegraphics[scale=0.45]{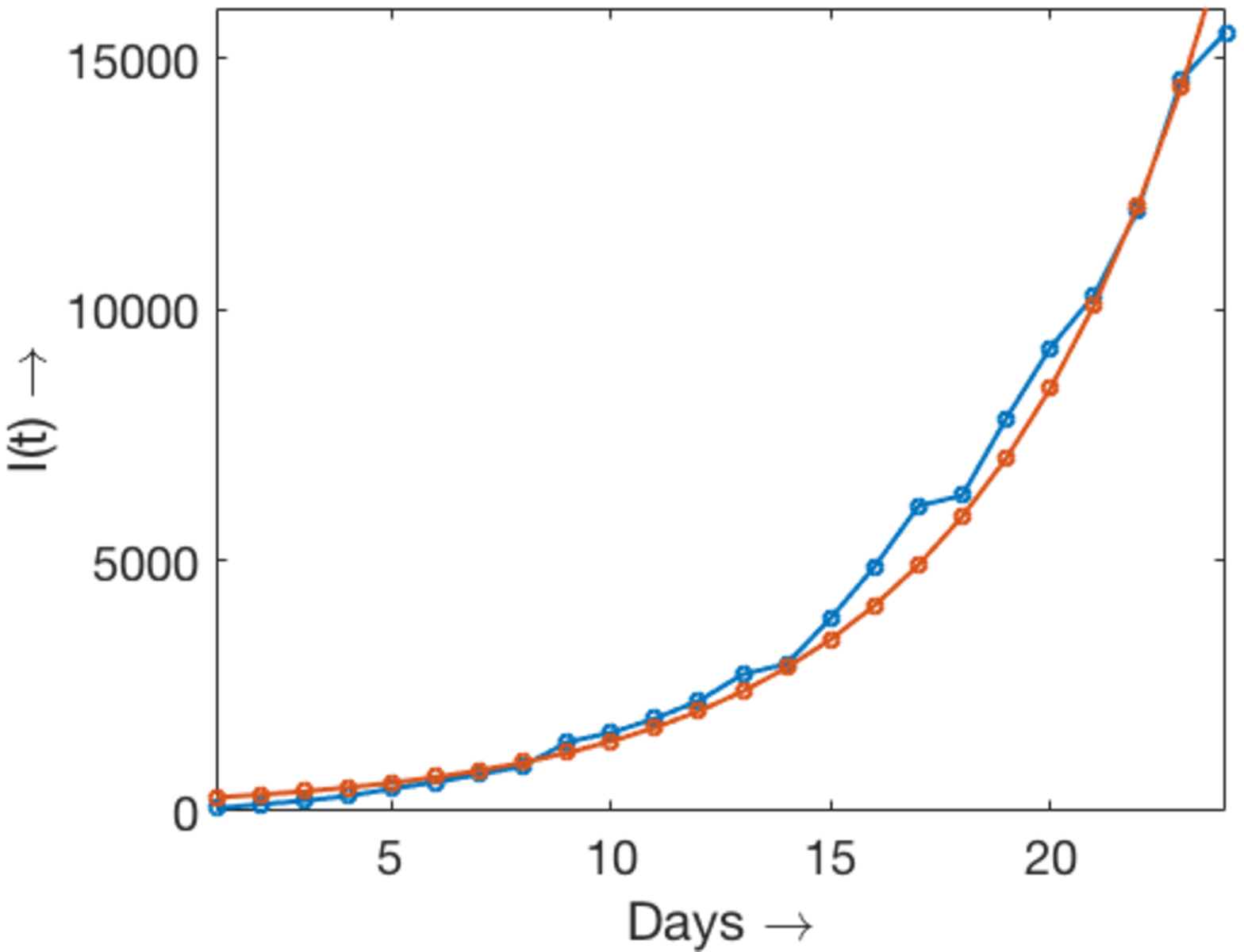}}
\caption{Latently infected individuals obtained as a sum of daily cases during the incubation period: (left) South Korea, the growing branch is approximated by the exponential $200 e^{0.21 x}$, decaying exponential by $35 \cdot 10^3 e^{-0.15x}$; (Right) Italy, the data curve is approximated by the exponential $230 e^{0.18 x}$. In both cases, the value $\tau=5$ days is used.}
\label{sum}
\end{figure}

Note that the data on daily reported cases do not exactly correspond to the variable $I(t)$. The latter represents a sum of daily cases during the incubation period. Taking into consideration that the incubation period to be of 5 days, we therefore obtain the graphs for the latently infected individuals. We observe growth-decay dynamics for South Korea (Figure \ref{sum}, left) and growth dynamics in Italy  (Figure \ref{sum}, right).

Fitting the data allows us to determine the basic reproduction number from equation (\ref{b15}). For South Korea,  $\mathcal{R}_0=1.123$ on the growing branch and $\mathcal{R}_0=0.932$ on the decaying branch. For Italy, $\mathcal{R}_0=1.103$. We mention here that the basic reproduction number depends on the incubation period.

\section{Discussion}

Our model confirms the efficiency of the approach to stop the disease spread by the limiting the number of contacts between the individuals through quarantine of infected individuals. This is quite obvious in theory, with the help of the simplest model formulation, but  difficult in practice. Success of the strategy also depends upon the appropriate time of implementation. Experience of China and South Korea shows that the peak of infection (maximum of newly reported cases on daily basis) is reached about 10 days after adopting serious restrictive measures. The number of infection increased during this time in 10-20 times.
In Italy 10 days after the universities and schools were closed (March 4) the peak of infection does not seem to be reached, and exponential growth continues.

\begin{figure}[ht!]
\centerline{\includegraphics[scale=0.3]{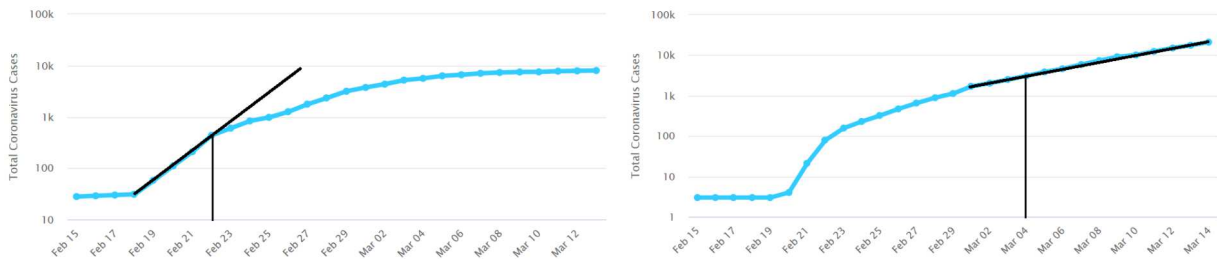}}
\caption{Total number of infected in the logarithmic scale.
In South Korea (top), the exponential growth rate (linear growth in this scale) before February 22 is replaced by a slower growth after this date. In Italy (bottom), the exponential growth rate (linear growth in this scale) after March 4 does not change.}
\label{kor-it}
\end{figure}

Moreover, the exponential growth rate of the total number of infected in China and in South Korea observed before the adopted measures (January 25 and February 22, respectively) rapidly changed to a slower growth rate afterward (Figure \ref{kor-it}). Similar situation is observed in Iran though the information about adopted measures is not fully available yet. However, in Italy the exponential growth rate does not change up to March 4. This can be an indication that the introduced measures are not sufficient or that they are not respected by local people.

\paragraph{Limitations of the model.}

The model has a number of obvious limitations. It takes into account only the beginning of the disease development where the number of susceptibles can be considered as constant. Arguably, it is a good approximation if the disease propagation is stopped/regulated when the number of infected is relatively small compared to the total population or the focus is on the early stages of the disease development.

Next, and perhaps mode importantly, the model does not take into account the spatial distribution of infected and their displacement. In this case, instead of equation (\ref{b3}) one has to consider a model with an explicit space, for instance the following equation:

\begin{equation}\label{b5}
  \frac{\partial I(x,t)}{\partial t} = \delta\frac{\partial^2 I(x,t)}{\partial x^2} + k I(x,t)S_0 - k I(x,t-\tau) S_0 ,
\end{equation}
where $I$ and $S_0$ now have the meaning of the corresponding densities insread of sizes and the diffusion term describes small-scale random motion of the individuals with intensity $\delta$ (neglecting long-distance jumps).

In the case where neither of $S_0$, $k$ and $\tau$ depend on space, Eq.~(\ref{b5}) can be reduced to the nonspatial model. Considering, for simplicity, the unbounded space (or a bounded interval with no-flux boundary conditions), we introduce the total size of infected as $J(t) = \int_{-\infty}^\infty I(x,t) dx$. Integrating equation (\ref{b5}) with respect to $x$, for variable $J$ we then obtain an equation similar to (\ref{b3}).

However, the spatially averaged model does not describe the dynamics in case some of the parameters depend on space. For instance, if there are two different patches of the disease development with different basic reproduction numbers, then the disease can be eradicated in the first patch due to the imposed restrictions but it can give a new outbreak in another patch if the restrictions are not adopted there or they are not sufficient. This situation is observed in Europe where the disease progresses exponentially  while it is already slowed down in China. Hence, in some cases, a multi-patch model should be considered with some connectivity between the patches at least for some time period.
 Also, the human movement can follow a pattern more complicated than is given by the Fickian diffusion, e.g.~to follow the network made by connections between large airports. 

Another important limitation of the model is related to the assumption of a single incubation period. According to the available data that are somewhat contradictory and far from complete, it is possible that there are different incubation periods or maybe the whole spectrum of incubation periods from several days up to four weeks. The distributed delay models are more appropriate in this case.

Finally, we mention virus mutations that can have a strong influence on the disease progression and treatment \cite{V}. At the moment, there are no available data on mutations of coronavirus, it will take some time before this aspect can be convincingly confirmed or ruled out.

\section*{Acknowledgements}

The first author acknowledges the IHES visiting program during which this work was done. 
The work was supported by the Ministry of Science and Education of Russian Federation, project number FSSF-2020-0018,
and by the French-Russian program PRC2307.

\end{document}